\documentclass[twocolumn,showpacs,amsmath,amssymb]{revtex4}

\usepackage{graphicx}
\usepackage{dcolumn}
\usepackage{bm}
\newcommand{\etal}{\emph{et al.}}
\newcommand{\be}{\begin{equation}}
\newcommand{\ee}{\end{equation}}
\newcommand{\bfig}{\begin{figure}}
\newcommand{\efig}{\end{figure}}
\newcommand{\incl}{\includegraphics}

\begin{document}

\title{Quantum interference in macroscopic crystals of 
non-metallic Bi$_2$Se$_3$}

\author{J. G. Checkelsky$^1$, Y. S. Hor$^2$, M.-H. Liu$^{1,\dagger}$, D.-X. Qu$^1$, R. J. Cava$^2$ and N. P. Ong$^1$
} 
\affiliation{Department of Physics$^1$ and Department of Chemistry$^2$,\\ 
Princeton University, New Jersey 08544, U.S.A.}

\date{\today}

\pacs{72.15.Rn,73.25.+i,71.70.Ej,03.65.Vf}

\begin{abstract}
Photoemission experiments have shown that Bi$_2$Se$_3$ 
is a topological insulator.  By controlled doping, we have obtained crystals of Bi$_2$Se$_3$ with non-metallic conduction.  At low temperatures, we uncover a
novel type of magnetofingerprint signal which involves the spin degrees of freedom. Given the mm-sized crystals, the observed amplitude is 200-500$\times$ larger than expected from universal conductance fluctuations. The results point to very long phase breaking lengths in an unusual conductance channel in these non-metallic samples.  We discuss the nature of the in-gap conducting states and their relation to the topological surface states.
\end{abstract}

\maketitle                   
A new class of insulators 
with non-trivial topological surface states has been predicted~\cite{Fu07,Moore07,Bernevig06,Bernevig06b,FuKane07}. 
The surface states of these topological insulators are 
chiral, and protected from disorder by a large
spin-orbit interaction that aligns the spin transverse to 
the wavevector. Angle-resolved photoemission spectroscopy (ARPES) has been
used to detect these surface states (SS) 
in Bi$_{1-x}$Sb$_x$~\cite{Hsieh08}. The spin polarization of the
SS in Sb was confirmed by spin-resolved ARPES~\cite{Hsieh09}. 
More recently, ARPES experiments showed that Bi$_2$Se$_3$~\cite{Xia09,Hsieh09b} and Bi$_2$Te$_3$~\cite{Shen09} are the simplest topological
insulators, with only a single Dirac surface state.

Due to its large band gap (300 mV), Bi$_2$Se$_3$ is  
a very attractive platform for exploring the transport properties
of the topological states. However, as-grown crystals of 
Bi$_2$Se$_3$ invariably display a metallic resistivity profile,
with the bulk chemical potential $\mu_b$ pinned to the conduction band.
By chemical doping, we have obtained
crystals of Bi$_2$Se$_3$ in which $\mu_b$ falls inside the gap. 
Although the residual conductance at low temperature $T$ (0.3 K)
appears to be still dominated by bulk conduction channels, we 
observe a novel type of conductance fluctuation phenomenon.
In a swept magnetic field $H$, the fluctuations retrace 
reproducibly, analogous to the magnetofingerprint signal 
reminiscent of universal conductance fluctuations (UCF)~\cite{LeeStone87}.  However, the fingerprint signal here
is observed in mm-sized bulk crystals rather than mesoscopic samples.  
Moreover, field-tilt 
experiments show that the carrier spin plays a role in generating the
flucutations. We discuss the highly unusual nature of the large
magnetofingerprint signal and possible connections to lattice dislocations
that bear topologically protected states.

\bfig[t]            
\incl[width=7cm]{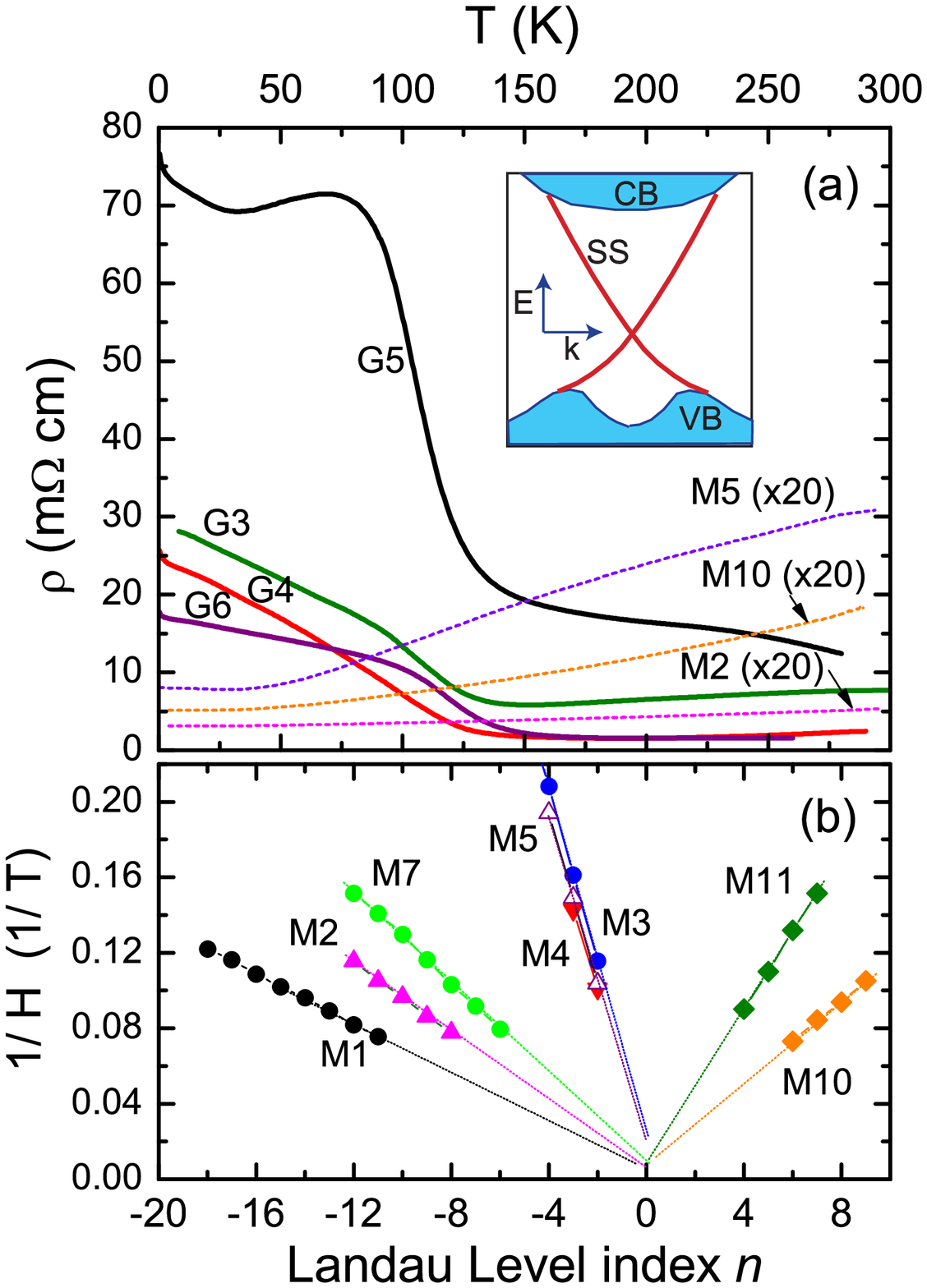} 
\caption{\label{figRT} (a)
Resistivity $\rho$ vs. $T$ in 4 samples (G3--G6) of Ca$_x$Bi$_{2-x}$Se$_3$
lightly doped with Ca to bring $\mu_b$ into the gap.  
Samples with $\mu_b$ not inside the gap (M2, M5 and M10)
display a metallic $T$ dependence (shown $\times$20).
The inset is a sketch of the surface states 
\cite{Xia09} crossing the gap from the VB to the CB.
(b) The LL index plot vs. field minima in the SdH oscillations 
observed in 8 metallic samples (M1$\cdots$M11). Negative (positive)
index $n$ represents the electron (hole) FS pocket.  As $\mu_b$ is 
lowered from the CB to VB, the FS area ${\cal S}_F\sim$1/slope
decreases before rising again.
} 
\efig

\bfig[t]            
\incl[width=8cm]{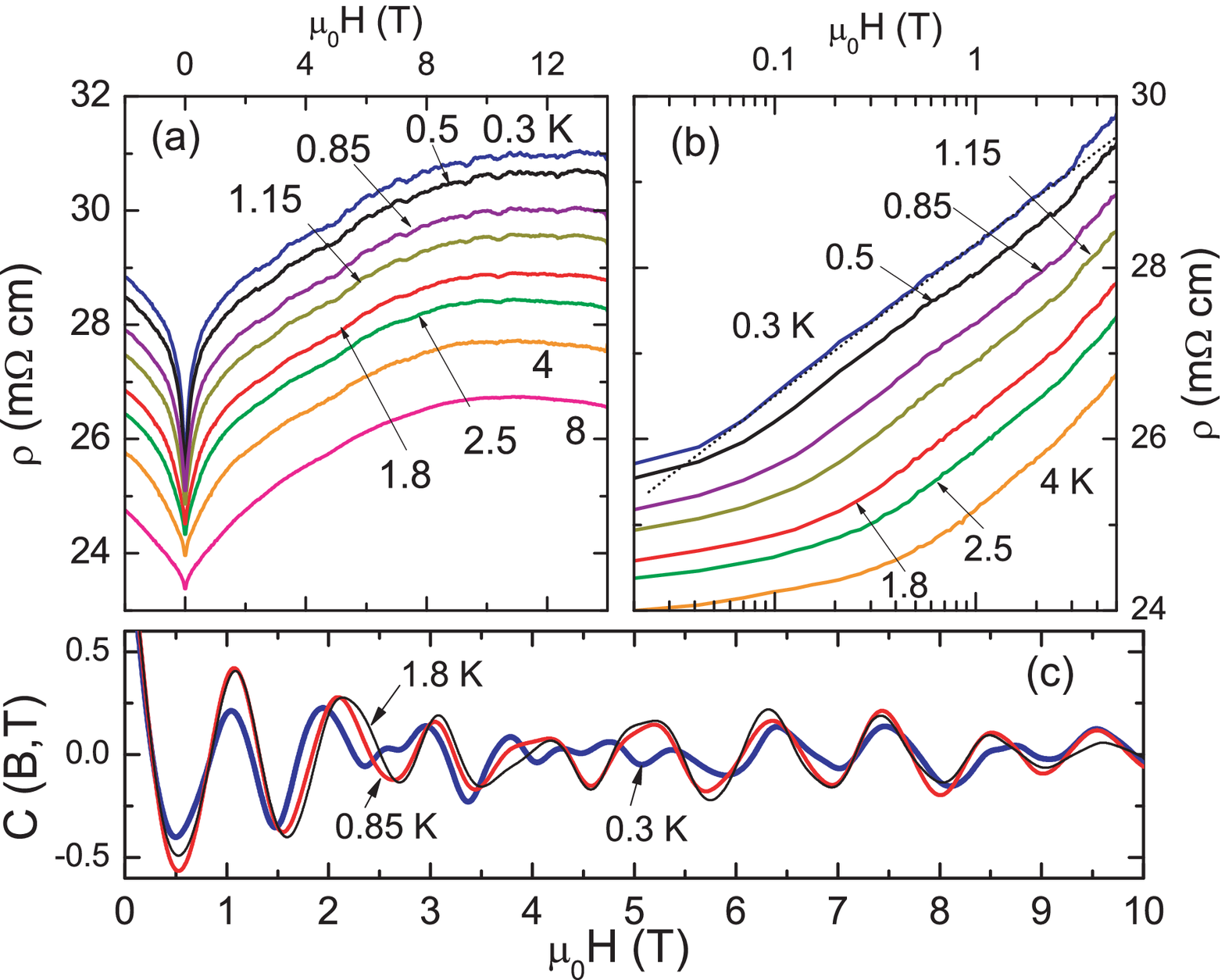} 
\caption{\label{figMR} 
Curves of $\rho$ vs. $H$ in Sample G4 at 0.3$<T<$8 K.  
The curves are plotted vs. $H$ (Panel a) and vs. $\ln H$
(b).  The MR displays a sharp anomaly of amplitude 10$\% R$ 
in weak $H$. In addition, large conductance fluctuations 
of amplitude up to 0.5$\%$ are resolved.  In Panel (b), 
the plot of $\rho$ vs. $\log H$ at 0.3 K shows that 
$\rho(H)\sim \ln H$ over 2 decades in $H$ (dashed line).
The slope is equivalent to $dG/d\ln H = 200 e^2/h$.
The correlation function ${\cal C}(B,T)$ of the fingerprint
signals ($\bf H||z$) is plotted in Panel (c) for $T$ = 0.3 K (bold curve), 
0.85 K (medium) and 1.8 K (thin). ${\cal C}(B)$ oscillates 
vs. $H$ instead of decaying as a power law.
} 
\efig

The ARPES results~\cite{Xia09} reveal that Bi$_2$Se$_3$ 
has a single Dirac surface 
state (SS) that crosses the bulk energy gap (Fig. \ref{figRT}a).  
In as-grown crystals, electrons donated by Se vacancies pin
$\mu_b$ to the conduction-band (CB) edge.  Recently, Hor \etal~showed that doping with Ca converts the crystals to $p$-type conductors~\cite{Hor09}.  
By tuning the Ca content $x$ in Ca$_x$Bi$_{2-x}$Se$_3$,
we progressively shift $\mu_b$ from the CB to inside the gap,
and then into the valence band (VB).   
Samples with $\mu_b$ in the CB or VB display Shubnikov de Haas (SdH) oscillations 
which we have used to measure the caliper 
area ${\cal S}_F$ of the bulk Fermi surface (FS).
The carrier sign was found by the thermopower and Hall effect.
The index plots of the Landau Levels (Fig. \ref{figRT}b) shows that ${\cal S}_F$ 
decreases (Samples M1$\to$M3) as $\mu_b$ enters the gap from the CB, and then
increases as $\mu_b$ exits the gap and moves into the VB.
The metallic profiles of $\rho$ are shown in expanded scale in Fig. \ref{figRT}a
(M2, M5 and M10). Using the changes in ${\cal S}_F$ to guide the doping, 
we have obtained non-metallic crystals in a narrow doping window 
0.002$< x<$0.0025, with $\mu_b$ lying inside the energy gap.  SdH oscillations
are not resolved in any of the non-metallic samples.

As plotted in Fig. \ref{figRT}a, $\rho$ of the non-metallic crystals (G3--G8) undergoes an increase to very large values (20--100 m$\Omega$cm) as $T$
falls below 130 K.  Below 20 K, $\rho$ approaches saturation 
instead of diverging as in a semiconductor.  Although the conductances $G$ of 
the non-metallic crystals are very poor (Table \ref{tab}), they are still 1000-8000$\times$ the universal conductance $e^2/h$ ($e$ is the charge and
$h$ is Planck's constant).  As we will describe, draining away
the high-mobility electrons in the CB uncovers a conductance channel 
of a highly unusual kind.

\begin{table}[b]
\begin{tabular}{|c|c|c|c|c|c|c|}		\hline
				& $\rho$	& $c$					 & $G$				&  rms$\delta G$ & $A$ & $n_H$ \\ \hline
units		& m$\Omega$cm & $\mu$m	&$e^2/h$	&$e^2/h$ 	& 	& $10^{18}$ cm$^{-3}$ \\  \hline\hline
G3			& 30			& 50					& -					&		-			&	- 	&	0.7 \\
G4			&	15			&	50    			&	8,000			&		5.9		&	178	& 5 \\		
G5			&	76			&	80					&	1,760			&		0.8		&	35	& 1	\\		
G6			&	18			&	50					&	7,000			&		1			&	135 & 7	\\		
G7			&	16			&	25					&	4,800			&		0.9		&	63	& 8	\\
G8			&	25			&	10					&	1,050			&		0.6		& 20	& 5	\\		\hline
\end{tabular}
\caption{\label{tab}
Sample parameters.  $c$ is the crystal thickness along $\bf c$.
Values of $G$ and rms$\delta G$ (in $e^2/h$) and $\rho$ 
are measured at 0.3 K (except for G3, which was not 
cooled below 4 K).  We define $A \equiv A_{orb}+ A_{spin}$ (see Eq. \ref{DG}).
The crystals are of nominal size 2 mm$\times$ 2 mm $\times$ c.
The Hall density $n_H= 1/eR_H$ is inferred from the Hall coefficient $R_H$.}
\end{table}

The magnetoresistance (MR) in Sample G4, measured
with $\bf H||\hat{c}||\hat{z}$ (with the current $\bf I||\hat{x}$), 
is displayed in Fig. \ref{figMR}a for $T$ = 0.3--8 K.  
In each curve, the most prominent
feature is the pronounced weak-field anomaly, which 
deepens to a sharp cusp at $H=0$.  Near the cusp, $\rho$ vs. $H$ 
follows a logarithmic behavior extending over 2 decades in $H$ at 0.3 K 
(Fig. \ref{figMR}b).  Moreover, at low $T$, large conductance fluctuations
($\sim 0.5\%\;\rho$) are apparent.  The retraceability of the
fluctuations versus $H$ distinguishes them from
random noise (the correlation function shown in Panel c is discussed later).

\bfig[t]            
\incl[width=9cm]{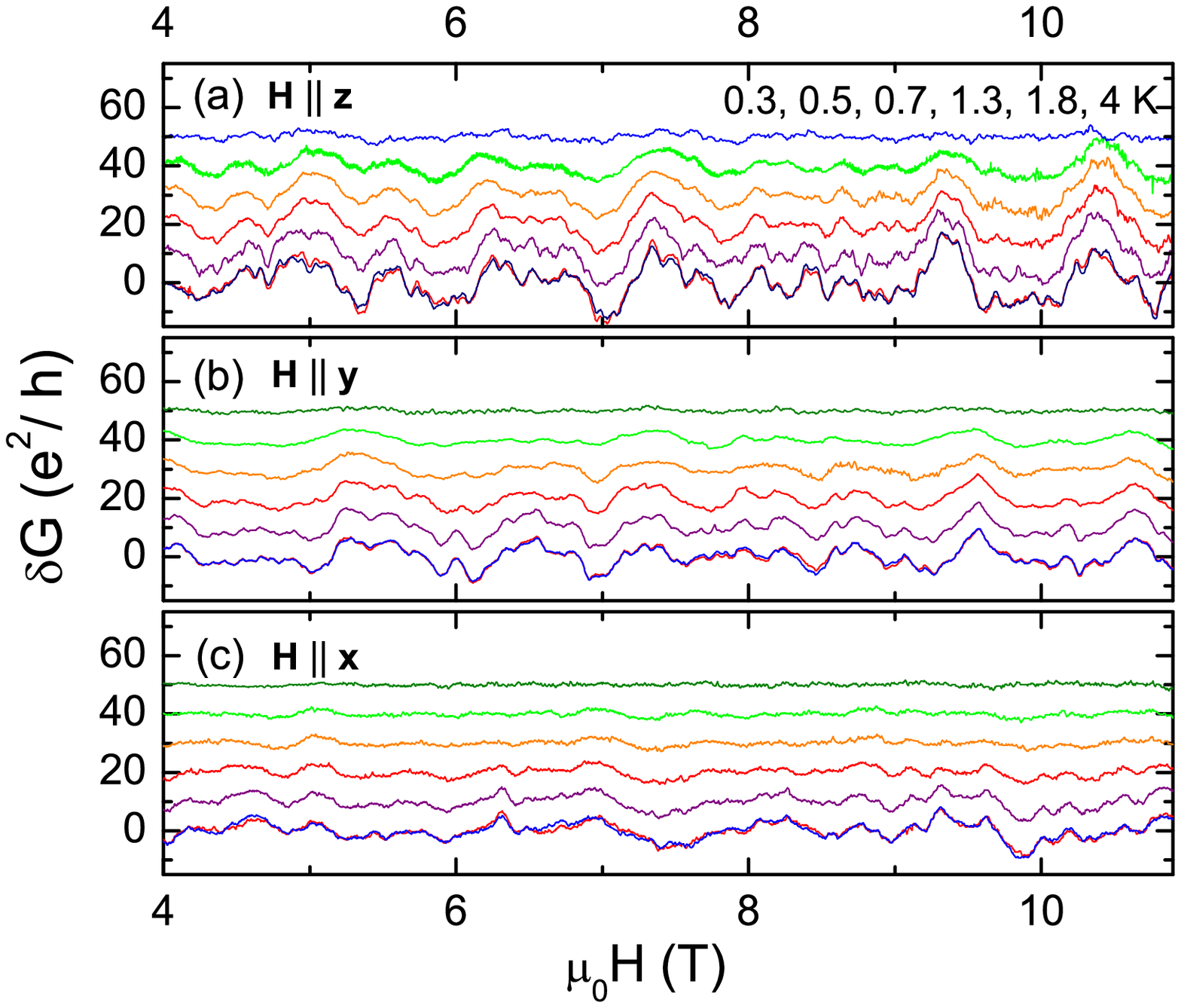} 
\caption{\label{figfluc} 
Curves of the magneto-fingerprint signal $\delta G(T,H)$ vs
$H$ in Sample G4.
The field $\bf H$ is aligned with $\bf \hat{z}$ (in Panel a), 
with $\bf\hat{y}$ (in b), and with $\bf\hat{x}||I$ (in c).
In each panel, curves are shown for five $T$ between 0.3 and 4 K 
(in ascending order). 
For clarity, adjacent curves are displaced vertically by 10 $e^2/h$.  
At 0.3 K, both up-sweep and down-sweep traces are
shown superposed to emphasize retraceability. 
}
\efig

Traces of the conductance fluctuations $\delta G(T,H) = G(T,H)-G_0(T,H)$, 
relative to the smoothed background $G_0(T,H)$, are displayed in 
Fig. \ref{figfluc} for 3 field orientations ($\bf H||\hat{z}$,
$\bf H||\hat{y}$ and $\bf H||\hat{x}$ in Panels a, b and c, respectively).  
In each panel, we have emphasized the 
retraceability of $\delta G(T,H)$ 
by superposing the up- and down-sweep curves at 0.3 K.  
Significantly,
they remain large even with $\bf H||I$ (Fig. \ref{figfluc}c).
We have followed the fluctuations to fields of 32 T in G4 and G8.  The
root-mean-square (rms) amplitude is nearly unchanged between 4 and 32 T (by contrast, SdH amplitudes
should grow exponentially).  In all non-metallic crystals 
studied at $T<$3 K (Table \ref{tab}), 
the magnetofingerprint is present.  

The existence of such large conductance fluctuations,
with amplitudes $\pm 10\;e^2/h$, is remarkable in a bulk mm-sized crystal
(2$\times$2$\times$0.05 mm$^3$ for G4).
Because magnetofingerprints imply field modulation of
the interference between conductance contributions from many channels, it
is natural to compare them with  
Aharanov Bohm (AB) oscillations and UCF investigated in mesoscopic
samples.  However, we will argue that they belong in a new category.

To see this, we recall the main features of the AB oscillations and UCF.
In mesoscopic, multiply-connected samples
(single loops or arrays of loops), two types of conductance oscillations
are observed with periods in flux $\phi = h/2e$ and $h/e$, respectively. 
The former, called AAS oscillations~\cite{Spivak81,Sharvin}, involves interference between a pair of time-reversed states that circumscribe a loop in opposite directions, and are observed only in very weak $H$ ($<$100 Oe).  The latter
(AB oscillations), arising from interference between waves that
traverse opposite arms of the loop, survive to very large $H$ ($>$10 T).  
However, the AB oscillations do not ensemble average.
When the sample size $L$ exceeds the phase-breaking length $L_{\phi}\sim$ 1 $\mu$m, the amplitudes decrease as $\sqrt{L_{\phi}/L}$~\cite{Umbach86}.  By contrast, the AAS oscillations have been observed -- with weak $H$ -- in giant arrays (10$^6$ loops) mm in size~\cite{Pannetier84}.

Apart from periodic oscillations, there exist weak, aperiodic fluctuations of the
conductance $G$ vs. $H$ (UCF) in simply-connected mesoscopic samples~\cite{Webb85,Chandrasekhar85}.  For electrons undergoing 
quantum diffusion in a phase-coherent region ($L\sim L_{\phi}$), the 
fluctuations $\delta G$ do not self-average. The rms value rms[$\delta G]$ is $\sim e^2/h$~\cite{Stone85,Altshuler,Lee85,LeeStone87}.  
Modulation of the interference by $H$ results in the magnetofingerprint trace.
The amplitude of UCF is also sharply suppressed if $L$ exceeds $L_{\phi}$.  In
metals, the phase-coherent region is actually cut off by
the thermal length $L_T= \sqrt{hD/k_BT}$, where $D$ is the 
diffusion constant.  As a result, UCF has been observed only in 1D (nanowires) and 2D systems (ultra thin-films and semiconductor devices) in samples 
with $L\le L_T\sim 1\mu$m~\cite{Webb85,Chandrasekhar85}.  
Magnetofingerprint signals in large $H$ have never been reported
in mm-sized samples (in 2D or 3D).

Hence, at $T\sim$ 1 K, both the AB oscillations and UCF, 
which persist to intense $H$, are confined to $\mu$m-sized samples, whereas
the AAS oscillations may be observed in mm-sized 
periodic arrays, provided $H<\sim$200 Oe.
These comparisons show that the fingerprint signal in Bi$_2$Se$_3$ is 
difficult to account for in terms of AB or AAS
oscillations or UCF. Specifically,
with $D\sim$ 0.05 m$^2$/s, we find $L_T\sim$ 1.5 $\mu$m at 1 K.  
The measured volume in G4 (2$\times 10^{-4}$ cm$^{3}$) exceeds 
the phase-coherent volume $L_T^3$ by a factor of 6$\times 10^7$.  By
classical averaging, the UCF should be strongly suppressed.
Quantitatively, the scaling of the variance of $G$ with $L$ is derived~\cite{LeeStone87} as $Var[G(L)] \sim (e^2/h)^2(L_T/L)^{4-d},$ with $d$ the dimension.  This gives the rms amplitude of UCF as 0.01-0.05 $e^2/h$, or 200-500 times weaker than depicted in Fig. \ref{figfluc}. 

Further insight into the fingerprint is obtained from its 
autocorrelation function 
\be
{\cal C}(B,T)= \frac{\langle \delta G(B',T)\delta G(B'+B,T)\rangle_{B'}}
{\langle \delta G^2\rangle},
\label{eq:corr}  
\ee
which measures how a particular peak of $\delta G$ is correlated 
with peaks at other field values ($\langle\cdots\rangle_{B'}$ means
averaging with respect to $B'$, and $B=\mu_0H$ with $\mu_0$ the permeability).  
As calculated~\cite{LeeStone87}, ${\cal C}(B)$ for UCF should have a power-law decay $\sim B^{d-4}$.  

In Fig. \ref{figMR}c, we plot the correlation ${\cal C}(B,T)$ calculated
from the results in Fig. \ref{figfluc}a.  Instead of a power-law decay, 
${\cal C}(B)$ is oscillatory, displaying $\sim$9 periods with an average 
period $B_p\sim$1 T.  If we interpret our period $B_p$ as a characteristic length 
$L_p\simeq \sqrt{\Phi_0/B_p}\;\simeq$ 640 \AA, the results
suggest a characteristic area $L_p^2$ in the network of conduction paths.

\bfig[t]            
\incl[width=8cm]{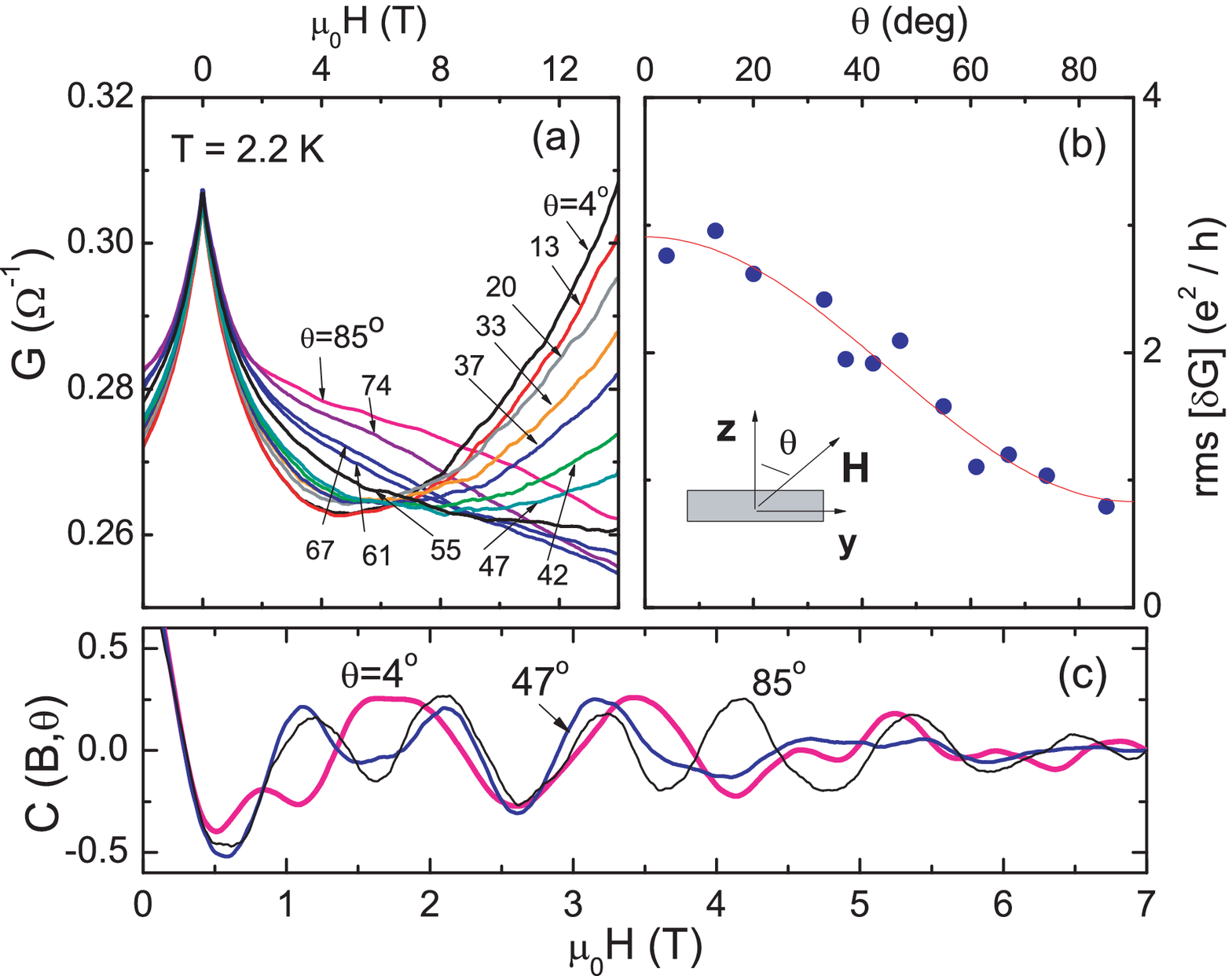} 
\caption{\label{figtheta} 
The magneto-fingerprint signal in tilted field.
(Panel a) The MR curves in G4 at selected field-tilt 
angles $4^{\rm o} <\theta<85^{\rm o}$ 
at $T$ = 2.2 K.  Panel (b) shows the rms amplitude
rms[$\delta G$] vs. tilt angle $\theta$
($\theta$ is defined in inset). The fit to 
rms[$\delta G$] = [$a + b\cos^2(\theta)](e^2/h)$ 
yields $a$ = 0.832 and $b$ = 2.08 (solid curve).
Panel (c) plots the correlation function
${\cal C}(B)$ vs. $H$ at selected $\theta$.
} 
\efig

The spin $\bf s$ of the carriers plays a significant role in the magnetofingerprint.  
We next describe the effect of tilting $\bf H$ at an angle $\theta$ to $\bf c$ 
in the $y$-$z$ plane.  Figure \ref{figtheta}a
shows plots of the conductance vs. $H$ in Sample G4 at 
selected $\theta$, with $T$ fixed at 2.2 K.  At each 
$\theta$, the curve is comprised of the weak-field quantum anomaly
superimposed on a parabolic (semi-classical) background.
At $\theta$ = 85$^{\rm o}$, the orbital term is negligible
compared with the spin term because the flux inside
the sample is reduced by a factor of 40.  Hence the anomaly
may be identified with the spin degrees alone.

As in Fig. \ref{figfluc}, we have extracted the conductance
fluctuations at each $\theta$.  The gradual decrease of the
amplitude rms[$\delta G$] vs. $\theta$ fits well to
$[a+b\cos^2\theta]e^2/h$ (Fig. \ref{figtheta}b).  The first term $a$ 
represents the spin term, while the second term ($\sim H_z^2$) is
from orbital coupling.
At maximum tilt (85$^{\rm o}$), only the spin term $a$ 
(comprising 29$\%$ of the rms) survives.  This is rare example of
a conductor in which the spin degrees are shown 
to generate a fingerprint signal.
The correlation function ${\cal C}(B,\theta)$
of the fluctuations (shown in Panel c) remains oscillatory 
with the same field period $B_p\sim$ 1 T at all $\theta$.

As displayed in Figs. \ref{figMR}a and b, the weak-field anomaly 
in the MR has a ln$H$ dependence that may be expressed as
\be
\Delta G(H) =  -\frac{e^2}{h} [ A_{orb}+ A_{spin}]\;\ln H,
\label{DG}
\ee
where $A_{orb}$ and $A_{spin}$ (the orbital and spin terms, respectively)
are both positive and comparable in magnitude.
A positive $A_{orb}$ implies field-suppresion 
of antilocalization in a 2D (two-dimensional) system (theory~\cite{Hikami} predicts $A_{orb}=1/2\pi$).  
In 2D systems, Coulomb interaction effects
lead to a spin-Zeeman term of the form in Eq. \ref{DG} with
$A_{spin} = \tilde{F}_{\sigma}/2\pi$, where the 
parameter $\tilde{F}_{\sigma}$ is of order 1~\cite{LeeRama85}. 
However, despite the suggestive
$\ln H$ dependence, we again encounter a large discrepancy in the
size of the anomaly. The fit in Fig. \ref{figMR}b 
yields $A_{orb}=122$ and $A_{spin}=78$ for G4.  
Both are 500-700$\times$ larger than the theoretical 2D values
(see Table I).

The most interesting question raised by these results concerns
the nature of the states in the gap responsible for the magnetofingerprint
signal in non-metallic Bi$_2$Se$_3$. Interpreting 
$G$ in G4 as a 2D sheet conductance would give $G/(e^2/h)= k_F\ell\sim$8,000,
which is far too large compared with $k_F\ell\sim$120 
given by $S_F$ measured in the metallic crystals.  This implies
that the in-gap states in G3-G8 are bulk-like.  
On the other hand, the log$H$ antilocalization anomaly identified 
in Eq. \ref{DG}) implies that they have 2D character
(despite the large $A_{orb}$ and $A_{spin}$).  We remark that 
there is evidence that ties the bulk in-gap states to the SS in Bi$_2$Se$_3$.  
In the APRES results, bulk states with 2D dispersion are seen to coexist with 
SS throughout the gap region (shown in yellow in 
Fig. 1a,b of Xia \etal~\cite{Xia09}).  
They propose that the bulk states are confined by
band-bending near the surface.  If hybridization 
with the SS is strong, we expect that the bulk states will also display a 
large Rashba coupling that locks the spin transverse to $\bf k$,
thereby explaining the role played by the spins.
If the 2D bulk gap states share the unusual
properties of the SS, the long-range phase coherent nature
of the fingerprint signal may be traced ultimately to the protected nature
of the SS.  The anomalies reported here provide a
window into the unusual transport properties of the SS.

We thank M. Feigel'man, L. Fu, M. Z. Hasan, C. L. Kane, P. A. Lee
and A. Yazdani for valuable discussions.
The research is supported by a MRSEC grant from the U.S. National Science Foundation (DMR 0819860).  Some of the results were obtained at
the National High Magnetic Field Laboratory at Tallahassee, a facility
supported by NSF, DOE and the State of Florida.


%

%
\end{document}